\documentclass[conference]{IEEEtran}
\normalsize
\usepackage[noadjust]{cite}
\usepackage{filecontents}
\usepackage{color,colortbl}
\usepackage[pdftex]{graphicx}
\usepackage{verbatim} 
\usepackage{epstopdf}
\usepackage[cmex10]{amsmath}
\usepackage[tight,footnotesize]{subfigure}
\usepackage{pifont}
\usepackage{algorithm}
\usepackage{algpascal}
\usepackage{algpseudocode}
\usepackage{multirow}
\usepackage{epstopdf}
\usepackage{amssymb}
\usepackage[colorlinks=true,pdfstartview=FitV,linkcolor=black,citecolor=black, urlcolor=blue,plainpages=false]{hyperref}
\usepackage{hyperref}
\definecolor{Gray}{gray}{0.99}
\newcommand{\E}{{\rm I\kern-.3em E}}

\usepackage{array}
\newcolumntype{P}[1]{>{\centering\arraybackslash}p{#1}}
\newcolumntype{M}[1]{>{\centering\arraybackslash}m{#1}}

\begin{document}

\title{Digital Predistortion in Large-Array Digital Beamforming Transmitters}

\author{\IEEEauthorblockN{Alberto Brihuega, Lauri Anttila, Mahmoud Abdelaziz and Mikko Valkama}
\IEEEauthorblockA{Tampere University of Technology, Department of Electronics and Communications Engineering,
Tampere, Finland}
\IEEEauthorblockA{Email: alberto.brihuegagarcia@tut.fi, lauri.anttila@tut.fi, mahmoud.abdelaziz@tut.fi, mikko.e.valkama@tut.fi}}

\maketitle

\begin{abstract}
In this article, we propose a novel digital predistortion (DPD) solution that allows to considerably reduce the complexity resulting from linearizing a set of power amplifiers (PAs) in single-user large-scale digital beamforming transmitters. In contrast to current state-of-the art solutions that assume a dedicated DPD per power amplifier, which is unfeasible in the context of large antenna arrays, the proposed solution only requires a single DPD in order to linearize an arbitrary number of power amplifiers. To this end, the proposed DPD predistorts the signal at the input of the digital precoder based on minimizing the nonlinear distortion of the combined signal at the intended receiver direction. This is a desirable feature, since the resulting emissions in other directions get partially diluted due to less coherent superposition. With this approach, only a single DPD is required, yielding great complexity and energy savings. 
\end{abstract}


\section{Introduction}

Large-scale antenna systems are one of the key technologies in future 5G systems, where the demands for higher data rates and network capacities have led the cellular network evolution towards utilizing higher frequency bands with hundreds of megahertz of available spectrum and deploying hundreds of antenna units at the base stations \cite{Intro_1}. In general, the energy efficiency of the networks is a very crucial factor \cite{Intro_1,Energy_efficiency, 5Green}, as the energy consumption of the ICT systems should preferably decrease. This is of particular importance in future large-array systems where high amounts of radio-frequency (RF) chains with power-hungry power amplifiers (PAs) and very wideband signals are deployed. 

In order to reduce the implementation and operating costs of future cellular networks, low-cost and energy-efficient RF components are expected to be utilized at the base stations \cite{GreenComm}. However, especially in case of PAs, high energy efficiency implies largely nonlinear operating characteristics. Such nonlinear hardware introduces then harmful distortion onto the transmit signal band, and more importantly, produces spectral regrowth of the transmitted waveform that leads to increased power leakage to the adjacent channels and might even violate the spurious emission limits \cite{Intro_6}.  In the context of antenna array transmitters, it is important to understand how  unwanted emissions behave in the spatial domain. In legacy single antenna transmitters, the out-of-band  (OOB) emissions exhibit the same spatial characteristics as those of the inband signal, and the amount of OOB radiated power is well defined by means of the adjacent channel leakage ratio (ACLR). However, in antenna array transmitters, new phenomena need to be considered. For instance, the signals can be radiated directionally, which may focus the nonlinear distortion in certain directions. In \cite{OOB_Mollen}, considering both line-of-sight (LoS) and non-line-of-sight (NLoS) propagation, it was shown that the ACLR, in the worst case scenario, is at the same level as in single antenna transmitters. Such worst case corresponds to the main beam direction, since it was shown that the OOB emissions also get coherently beamformed regardless of LoS or NLoS propagation, while in the other directions, the nonlinear distortion gets diluted due to non-coherent superposition. The spatial domain is thus the key to understand how unwanted emissions behave in array transmitters and to develop more efficient solutions towards their linearization.

There are many different approaches that allow to reduce the above-mentioned nonlinear distortion. For instance, applying a back-off to the power amplifier input signal, such that the signal does not essentially span to the nonlinear operating region of the PA, is an easy but unattractive approach since it requires using larger PAs operating with low power efficiency. In digital predistortion (DPD), a nonlinear block is inserted prior to the power amplifier stage, to compensate for its nonlinear behaviour. DPD is a more attractive technique compared to back-off, since it enables more efficient linear operation of the PAs. In general, predistortion is implemented in the digital domain, and a dedicated predistorter per RF chain is utilized.

DPD methods in the context of array transmitters have been studied in the very recent literature to a certain extent. In \cite{DPD_MM_1}, digital predistortion is addressed assuming fully digital beamforming transmitters, primarily focusing on the reduction of the complexity of the learning algorithm. A dedicated DPD unit per RF chain is considered, implying that there are as many predistorter blocks as antenna units, which may not be a desirable solution for large-array transmitters.
DPD processing in single-user hybrid MIMO context was investigated in \cite{HybridMIMO,DPD_MM_3,DPD_MM_2,DPD_MM_4,DPD_HMIMO}. In \cite{DPD_MM_3}, it is assumed that all the power amplifiers within the transmitter are identical, while in \cite{DPD_MM_2}, the DPD solution was devised based on observing only one of the PAs. As a result, both approaches lead to reduced linearization performance due to differences between the characteristics of real power amplifiers. In \cite{HybridMIMO}, the authors proposed a novel and efficient solution for linearizing a set of PAs within an antenna subarray under pure LoS propagation, relying on the fact that OOB emissions are more significant in the main beam direction. Therefore, by coherently combining the PA output signals within the subarray, it is possible to mimic the signal received by the intended user and the DPD results in minimizing the nonlinear distortion towards the main beam direction. In other spatial directions it is the combined effect of the DPD and array beam pattern what keeps the OOB emissions at a sufficiently low level.

In this article, we propose a novel DPD approach and parameter learning architecture in the context of single-user fully digital beamforming transmitters, which are much simpler than current state-of-the-art techniques that assume a dedicated DPD block per power amplifier. Specifically, we propose to perform the predistortion prior to the baseband (BB) precoding block, at data stream level, requiring thus only a single DPD to linearize an arbitrarily large set of power amplifiers. Based on the fact that most of the unwanted emissions take place in the direction of the intended receiver, the purpose is to minimize the emissions in this direction. In order to do so, a replica of the received signal at the intended receiver is pursued and calculated, assuming that the channel state information and the estimates of the direct models of the power amplifiers are available. This replica is then utilized to perform the DPD learning, which follows a decorrelation-based learning rule, similar to \cite{Dec_Based}. Then, we also propose an alternative learning architecture that allows to further reduce the DPD parameter estimation, while still building on the structure of the nonlinear distortion at the intended receiver.

The rest of this paper is organized as follows: In Section \ref{sec:system_modeling}, the system model is described and basic modeling of PA nonlinear distortion in digital beamforming array transmitter system is provided. Then, in Section \ref{sec:dpd_modeling}, the proposed DPD structure and the parameter learning solution are introduced and described. In Section \ref{sec:complexity}, an analysis of the complexity of the proposed DPD solution and its comparison against the current state-of-the-art solutions are provided.  Then in Section \ref{sec:results}, the numerical performance evaluation results are presented and comprehensively analyzed. Lastly, Section \ref{sec:conclusion} will provide the main concluding remarks.

\section{System Model and Nonlinear Distortion in Digital MIMO Transmitters}\label{sec:system_modeling}

\begin{figure}[t!]
    \centering
    \includegraphics[width=1\linewidth]{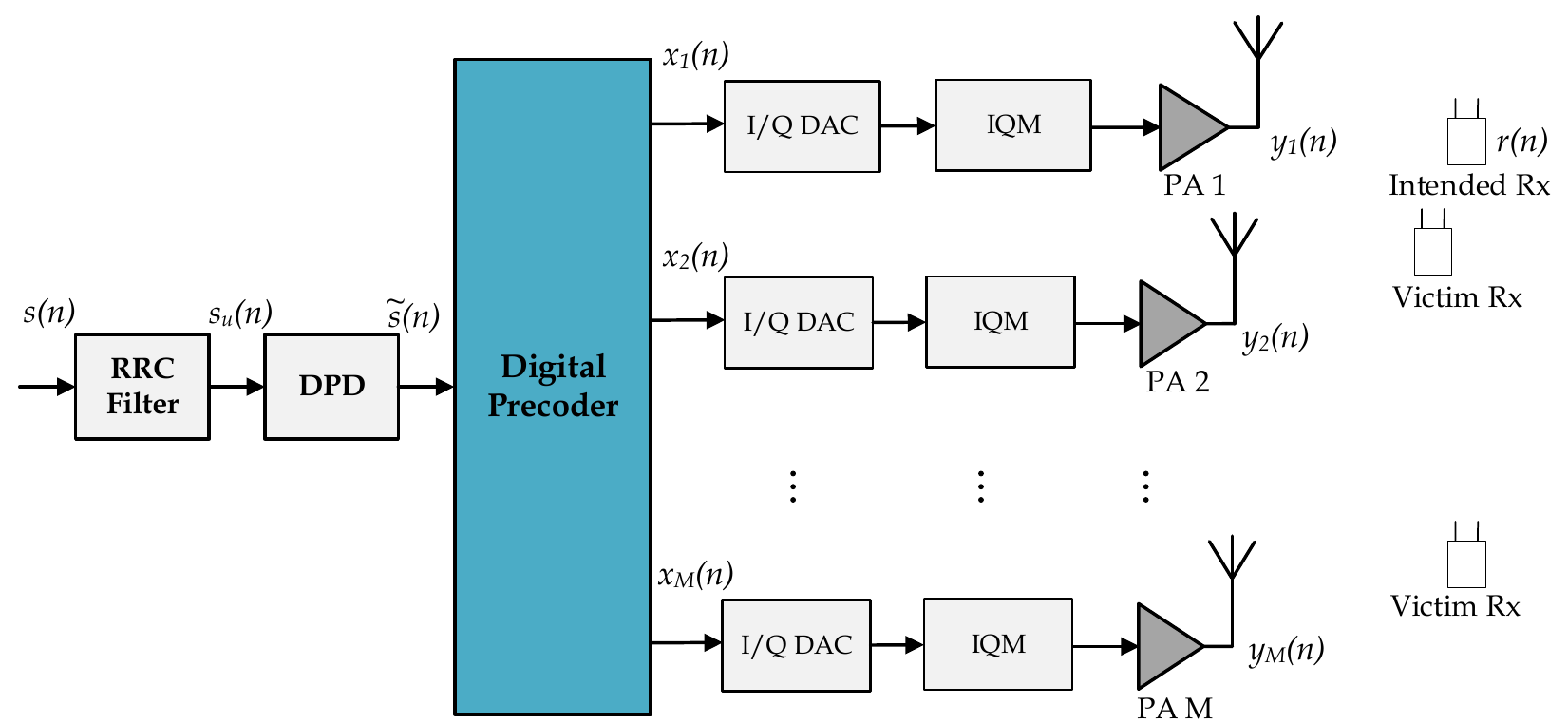}
    \caption{Considered system with digital beamforming based large-array transmitter, where the signal is radiated towards the intended user's direction. Potential victim receivers utilizing the adjacent channel that are sensitive to the OOB radiation produced by the nonlinear PAs are also shown.}
    \label{fig:architecture}
\end{figure}
In this section, basic mathematical modeling of the nonlinear distortion is pursued, with particular emphasis on the combined received signal. We assume a single-user large-scale digital beamforming transmitter with single-carrier transmission with the system architecture as depicted in Fig. \ref{fig:architecture}. $M$ stands for the number of transmit antennas available at the base station and $s(n)$ denotes the data stream intended for the receiver. Upsampling and filtering with a root-raised cosine (RRC) pulse-shape yields the DPD input signal $s_u(n)$. In the basic system modeling, the DPD block is assumed to be turned-off, while is then explicitly considered later in the paper. It is also important to note that the RRC filtering and the DPD processing both take place prior to the baseband (BB) digital precoding stage, and therefore, only a single DPD and a single RRC filter are required. The upsampled and filtered data stream is then spatially precoded by means of a phase-only-matched-filter precoder, that applies phase rotations to the precoder input such that the transmitted signals are combining coherently at the intended receiver. For simplicity, we consider only phase rotations in the digital precoder since, otherwise, the combined nonlinear distortion would essentially depend on the exact channel state and hence the DPD would need to be updated within the coherence time of the channel. The digital precoder is denoted by $\mathbf{w} = [w_1, w_2,\dots, w_M]^T$, and the precoded samples are obtained as $\mathbf{x}(n) = \mathbf{w}s_u(n) = [x_{1}(n), x_{2}(n), \dots, x_{M}(n)]^{T}$, where $x_{m}(n)$ stands for the signal at the m-th antenna branch. For mathematical tractability, we assume narrowband transmission and therefore, memoryless power amplifiers and memoryless channel models are considered in this work. Extensions to more sophisticated memory polynomial type PA models and frequency-selective channels are considered in our future work. 

The $m$-th PA output signal assuming $P$-th order memoryless polynomial models, and yet without DPD processing, reads
\begin{align}\label{PA_model}
    y_m(n) &= \sum_{\substack{p=1 \\ p, \text{odd}}}^{P} \alpha_{p,m} x_m(n) |x_m(n)|^{p-1} \\
     &= \sum_{\substack{p=1 \\ p, \text{odd}}}^{P} \alpha_{p,m} w_m |w_ms_u(n)|^{p-1}s_u(n), \label{PA_output}
\end{align}
where $w_m$ denotes the BB precoder coefficient corresponding to the $m$-th antenna branch, while $\alpha_{p,m},p=1,3,...,P$ denote the corresponding PA coefficients. Considering now that the precoder entries have unit modulus, i.e., $|w_m| = 1$, (\ref{PA_output}) can be re-written as
\begin{equation}
y_m(n) = w_m\sum_{\substack{p=1 \\ p, \text{odd}}}^{P} \alpha_{p,m} |s_u(n)|^{p-1}s_u(n). \label{linear_phase_rotation}
\end{equation}
Then, if we let $\mathbf{h} = [h_1, h_2,\dots, h_M]^T$ denote the zero-mean-unit-variance flat-fading Rayleigh spatial channel vector, the received signal at the intended receiver reads
\begin{align}
    r(n) &= \sum_{\substack{m=1}}^{M} h_my_m(n) \\
    &= \sum_{\substack{m=1}}^{M}\sum_{\substack{p=1 \\ p, \text{odd}}}^{P}\alpha_{p,m} h_m w_m|s_u(n)|^{p-1}s_u(n), \label{received_signal}
\end{align}
where $y_m(n)$ is the signal at the $m$-th transmit antenna port, while $h_m$ stands for its corresponding channel coefficient. For simplicity, we have excluded additive channel noise from the above received signal model to focus on the essentials. From (\ref{received_signal}) it can be seen that the nonlinear distortion observed by the intended user is a weighted linear combination of the static nonlinear (SNL) basis functions of the form $u_p(n) = |s_u(n)|^{p-1}s_u(n)$, $p=1,3,...,P$. 


\section{Proposed DPD Structure and Parameter Learning Solution}\label{sec:dpd_modeling}
\begin{figure}[t!]
    \centering
    \includegraphics[width=1\linewidth]{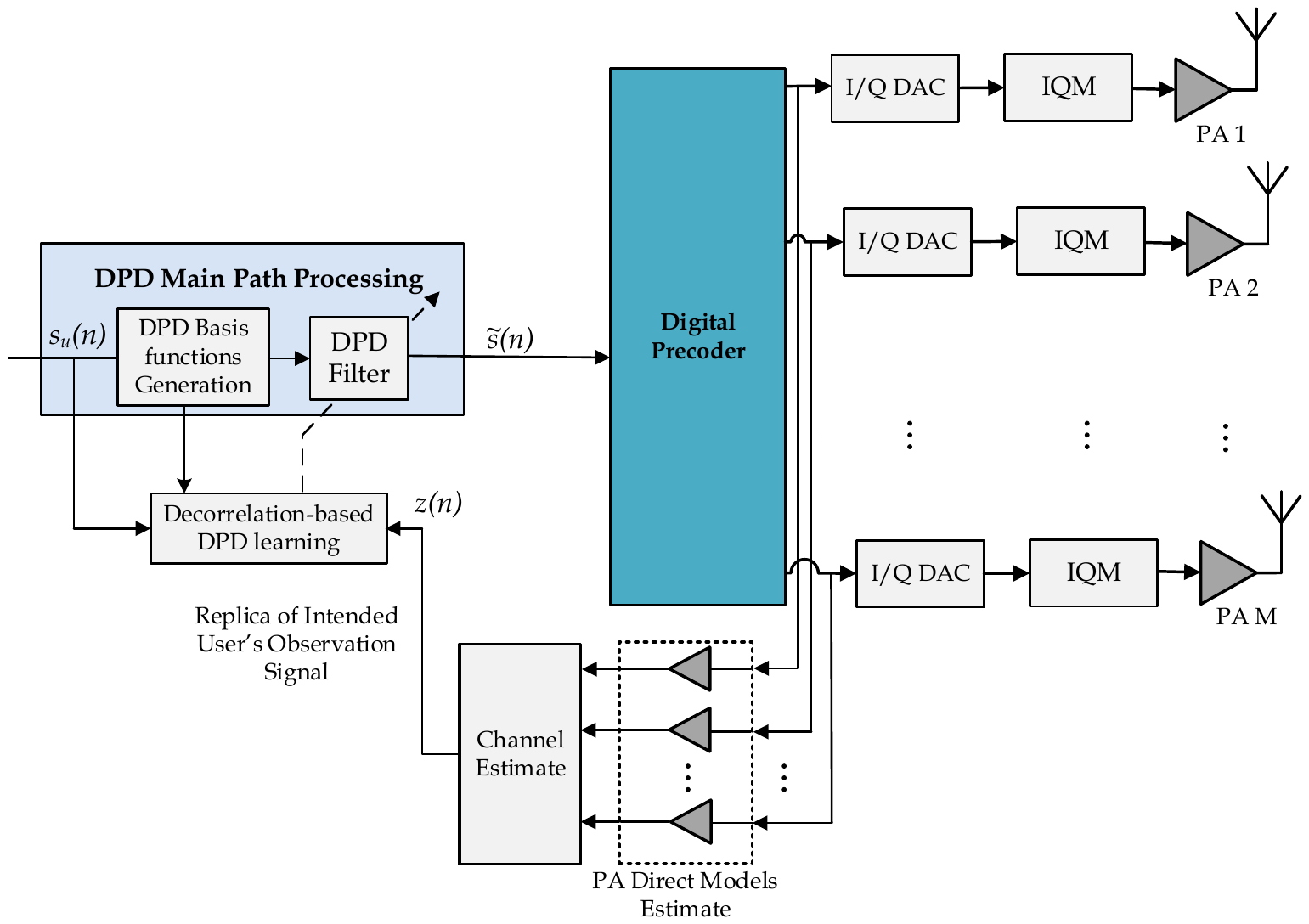}
    \caption{DPD learning architecture where the precoder output signals together with the estimates of the PA direct models and the channel state information are used to generate a replica of the signal observed by the intended receiver, which is used for DPD parameter learning.}
    \label{fig:DPD_architecture}
\end{figure}

In this section, we introduce the DPD architecture and parameter learning solution. The proposed architecture is depicted in Fig. \ref{fig:DPD_architecture}, where a single DPD unit is utilized to linearize a set of $M$ parallel power amplifiers. Since the characteristics of the $M$ involved PAs are in general different, the linearization task is challenging, basically resulting in an underdetermined problem. However, by focusing on the combined received signal, a well-defined problem is obtained.

\subsection{Proposed DPD Structure}

Based on the received signal model derived in Section \ref{sec:system_modeling}, we now describe the proposed DPD architecture. 
We first express (\ref{received_signal}) such that the linear and nonlinear terms are separated as follows
\begin{equation}\label{su_rx_dist}
    \begin{split}
    r(n) &=  \sum_{\substack{m=1}}^{M}\alpha_{1,m} h_m  w_m u_1(n)\\
    &+ \sum_{\substack{m=1}}^{M} \sum_{\substack{p=3 \\ p, \text{odd}}}^{P} \alpha_{p,m} h_m w_m u_p(n).
    \end{split}
\end{equation}
 In the DPD processing, we focus our attention only on the nonlinear terms in (\ref{su_rx_dist}), since the linear term behaves similarly as in any ordinary linear communication system, and can thus be properly processed and equalized at the receiver. The main idea in the DPD processing is to generate an appropriate low-power injection signal, with similar structure to the SNL basis functions, such that the nonlinear terms in (\ref{su_rx_dist}) are minimized at the receiver. 
 This injection signal is obtained by utilizing the above described nonlinear basis functions and a proper set of DPD coefficients, denoted by $\beta_q$. The predistorter output signal thus reads \cite{Concurrent_abdelaziz}
\begin{equation}
    \tilde{s}(n) = s_u(n) + \sum_{\substack{q=3 \\ p, \text{odd}}}^{Q}\beta_q^*u_q(n),
    \label{injection}
\end{equation}
where $\beta_q$ is the $q$-th order DPD coefficient and $Q$ denotes the DPD order. We note that the DPD injection signal, described in the second term of (\ref{injection}), is in general a low power signal that can be assumed to essentially only excite the linear response of the power amplifier. Thus, the weak higher order nonlinear terms that result from the cascade of the DPD and the PA nonlinearities are neglected in the following analysis for mathematical tractability.

Next we explicitly demonstrate how the above injection signal principle allows to cancel the combined nonlinear distortion at the receiver. By taking into account the predistorted signal $\tilde{s}(n)$, and considering that the precoder coefficients are chosen following the phase-only matched filter principle, i.e., $w_m = e^{-j\angle h_m}$, the received signal (\ref{su_rx_dist}) can be re-written as 
\begin{equation}\label{RX_DPD}
\begin{split}
   \tilde{r}(n) &= \sum_{\substack{m=1}}^{M}\alpha_{1,m} |h_m|\tilde{s}(n) \\
    &+ \sum_{\substack{m=1}}^{M} \sum_{\substack{p=3 \\ p, \text{odd}}}^{P} \alpha_{p,m} |h_m||\tilde{s}(n)|^{p-1}\tilde{s}(n). 
\end{split}
\end{equation}
Substituting now (\ref{injection}) into (\ref{RX_DPD}), and utilizing the low-power assumption for the injection signal, we obtain
\begin{equation} \label{rx_withDPD}
    \begin{split}
    \tilde{r}(n) &= \sum_{\substack{m=1}}^{M}\alpha_{1,m} |h_m|u_1(n) \\
         & + \sum_{\substack{m=1}}^{M} \sum_{\substack{p=3 \\ p, \text{odd}}}^{P} \alpha_{p,m} |h_m|u_p(n) \\ 
         &+ \sum_{\substack{m=1}}^{M} \sum_{\substack{q=3 \\ p, \text{odd}}}^{Q}\alpha_{1,m} \beta^*_{q} |h_m|u_q(n).
    \end{split}
\end{equation}
where the SNL basis function notation of the form $u_p(n) = |s_u(n)|^{p-1}s_u(n)$ is adopted.

Now, if the DPD order $Q$, and the PA nonlinearity order $P$ are assumed to be equal, for simplicity, (\ref{rx_withDPD}) can be re-written as
\begin{align}
   \nonumber \tilde{r}(n) &= \sum_{\substack{m=1}}^{M}\alpha_{1,m} |h_m|u_1(n) \\ 
    &+ \sum_{\substack{m=1}}^{M} \sum_{\substack{p=3 \\ p, \text{odd}}}^{P}(\alpha_{1,m} \beta^*_{p} |h_m| + \alpha_{p,m} |h_m|)u_p(n)\\
 &=\alpha_{1,tot} u_1(n) + \sum_{\substack{p=3 \\ p, \text{odd}}}^{P}\left(\beta^*_p\alpha_{1,tot}+\alpha_{p,tot}\right)u_p(n), \label{rx_final}
\end{align}
where $\alpha_{p,tot} = \sum_{\substack{m=1}}^{M}\alpha_{p,m} |h_m|$, $p=1,3,...,P$. From (\ref{rx_final}), it is possible to observe that the DPD coefficients $\beta^*_p$ can be selected such that the nonlinear terms at the receiver end are completely cancelled, and hence the DPD approach allows for efficiently minimizing the combined nonlinear distortion at the receiver.
\subsection{Combined Feedback Architecture and DPD Learning}
The proposed feedback signal and parameter learning architecture are depicted in Fig. \ref{fig:DPD_architecture}. The fundamental idea is to generate a local replica of the intended user's received signal 
at the transmitter side, such that it can be used to learn the DPD coefficients. The learning architecture is thus essentially mimicking the true transmission and over-the-air combining of the transmit signals towards the intended receiver. The DPD coefficients are obtained by means of a decorrelation-based learning rule that targets to minimize the correlation between the local replica of the feedback signal, denoted by $z(n)$ in the continuation, and the SNL basis functions. Such DPD learning approach was first introduced by the authors in \cite{Concurrent_abdelaziz}, in the context of single antenna transmitters, while is here utilized in array transmitter context. In order to calculate the feedback signal, we utilize the precoder output signals together with the estimates of the PA direct models that introduce the nonlinear distortion onto the feedback signal in a similar manner as the true PAs in the actual transmission. The channel estimates reproduce then the combined signal at the intended receiver emulating the true propagation. Consequently, the feedback observation signal, denoted by $z(n)$, reads
\begin{align}
    z(n) &= \sum_{\substack{m=1}}^{M}\alpha^{e}_{1,m} |h^e_m|u_1(n) \nonumber\\
    &+ \sum_{\substack{m=1}}^{M} \sum_{\substack{p=3 \\ p, \text{odd}}}^{P} \alpha^{e}_{p,m} |h^e_m|u_p(n) \\
    &= \alpha^e_{1,tot}u_1(n) + \sum_{\substack{p=3 \\ p, \text{odd}}}^{P} \alpha^e_{p,tot}u_p(n), \label{observation_signal}
\end{align}
where $|h^e_m|$ stands for the channel estimate, $\alpha^{e}_{p,m}$, $p=1, 3, ..., P$ refer to the estimated coefficients of the direct model of the $m$-th PA while $\alpha^e_{p,tot} = \sum_{\substack{m=1}}^{M}\alpha^e_{p,m} |h^e_m|$. The feedback observation signal in (\ref{observation_signal}) has thus the same structure as the true received signal in (\ref{rx_final}). The observation signal is then utilized to generate the error signal considered for learning the DPD coefficients, expressed as
\begin{equation} \label{error signal}
    e(n) = z(n) - Gs_u(n),
\end{equation}
where $G$ is the effective or combined complex linear gain of the feedback observation path. For further details on how to perform the coefficient learning based on the decorrelation-based approach, with the help of the error signal in (\ref{error signal}) as well as the SNL basis function samples, please refer to \cite{Concurrent_abdelaziz}.

\begin{figure}[t!]
    \centering
    \includegraphics[width=1\linewidth]{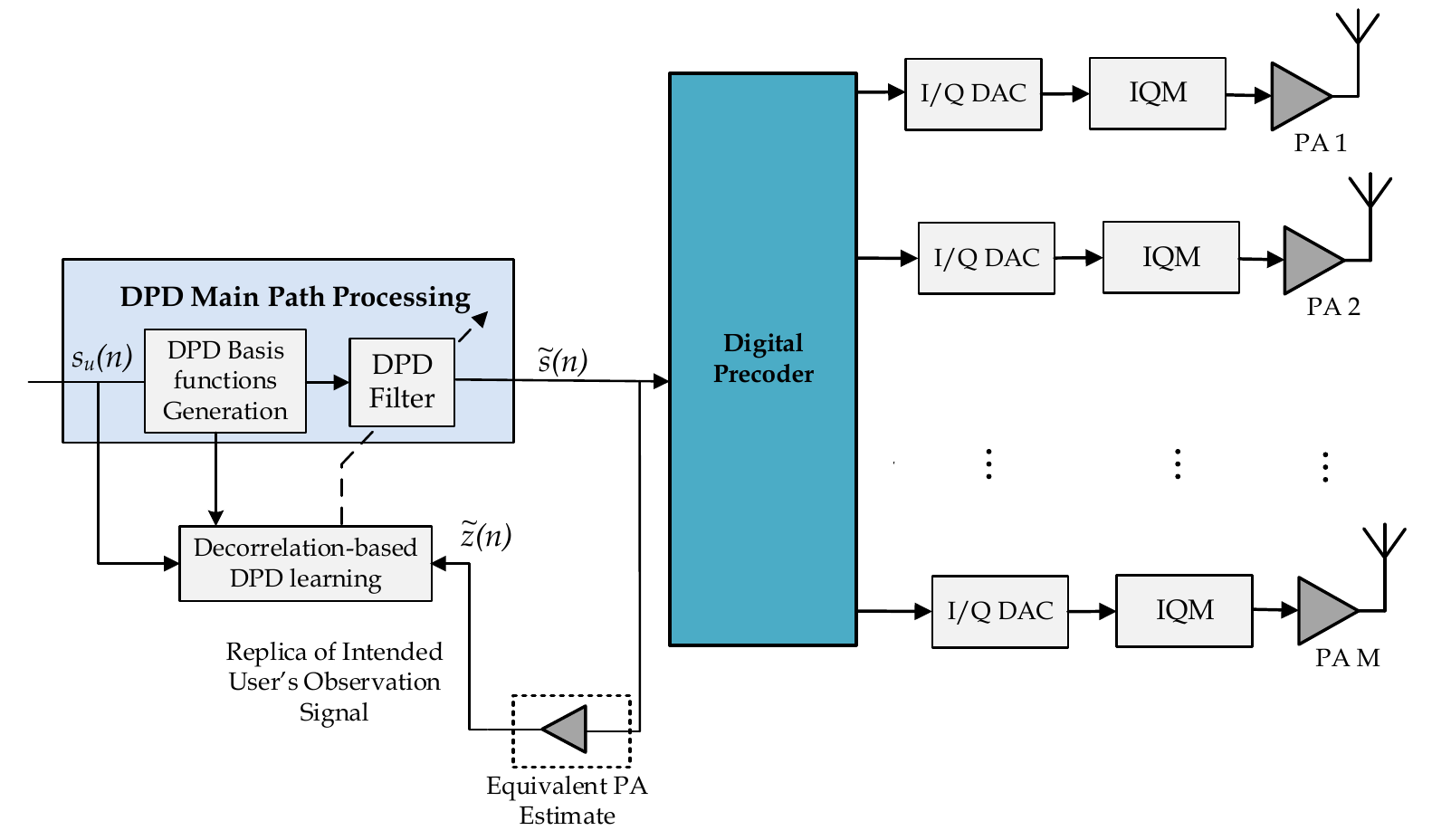}
    \caption{Reduced complexity DPD learning architecture where the unprecoded data stream and the equivalent PA estimate of the array are used to generate an approximate replica of the signal observed by the intended receiver. 
    }
    \label{fig:DPD_simplified}
\end{figure}

\subsection{Simplified Learning Architecture}
Interestingly, equations (\ref{rx_final}) and (\ref{observation_signal}) indicate that it is possible to derive a simplified learning architecture allowing for a further reduction in the complexity. The received signal model in (\ref{rx_final}) is a weighted linear combination of the user's data stream and the SNL basis functions, with weights that depend on the PA coefficients and the channel responses. 
The DPD coefficients are thus essentially learned based upon the effective combined coefficients $\alpha_{1,tot}$ and $\alpha_{p,tot}$. These effective coefficients can be interpreted as the equivalent nonlinear array response of the precoder, PAs and the channel, thus there is no need for mimicking all the $M$ parallel PAs within the transmitter individually, only this effective combined response. In addition, the BB precoder only introduces phase rotations that do not affect the nonlinear behavior of the transmitter as shown in (\ref{linear_phase_rotation}) already. Hence, modeling wise, one can consider feeding the same unprecoded data stream samples in parallel to all PA models and coherently combining the outputs. Lastly, to strive for complexity reduction in the learning loop, we drop the exact channel estimates in terms of the amplitudes. This yields, strictly speaking, only an approximation of the feedback signal in (\ref{observation_signal}), however, as it will be shown through numerical examples it does not have any essential impact on the linearization performance. The feedback signal for the reduced complexity learning architecture is thus given by
\begin{table*}[t!]
\caption{DPD computational complexity quantification and comparison. Proposed architecture and Reduced complexity refer to the DPD solution proposed in this work and its reduced complexity version, respectively.}
\centering
\begin{tabular}{|p{0.26\linewidth}|M{0.245\linewidth}|M{0.245\linewidth}|M{0.145\linewidth}|M{0.13\linewidth}}
\hline
 & Proposed Architecture & Reduced Complexity & Reference Solution \\ 
\hline
Upsampling and Filtering (FLOP/sample) &$2(2L-U)$ &$2(2L-U)$ & $2(2L-U)M$  \\

BF generation (FLOP/sample)  & $Q+2$ & $Q+2$ & $M(Q+2)$ \\

DPD main processing (FLOP/sample)  & $4(Q-1)$ & $4(Q-1)$ &$4M(Q-1)$ \\

Spatial Precoder (FLOP/sample) & $6MU$ & $6MU$ & $6M$  \\

\textbf{Transmission Complexity (GFLOP)} & \textbf{0.131} & \textbf{0.131} & \textbf{0.528}  \\
\hline
BF Orthogonalization (FLOP/N samples)  & $4(N+5/3)5^2$ & $4(N+5/3)5^2$  & $4M(N+5/3)5^2$ \\

PA Estimation (FLOP/N samples) &$4M(N+5/3)5^2$ &$4(N+5/3)5^2$ & $0$  \\

DPD learning (FLOP/KN samples)  & $ 92KN +  MNK(Q+54)  + 2(Q-1)K$ & $ 92KN +  NK(Q+40)  + 2(Q-1)K$&  $MK[92N + 2(Q-1)]$\\  

\textbf{Learning Complexity (GFLOP)} &\textbf{4.546} & \textbf{0.302} & \textbf{6.28}  \\
\hline
\end{tabular}\label{Tab:Table1}
\end{table*}

\begin{align}
    \tilde{z}(n) = \tilde{\alpha}^e_{1,tot}u_1(n) + \sum_{\substack{p=3 \\ p, \text{odd}}}^{P} \tilde{\alpha}^e_{p,tot}u_p(n), \label{observation_signal_red}
\end{align}
where $\tilde{\alpha}^e_{p,tot} = \sum_{\substack{m=1}}^{M}{\alpha}^e_{p,m}$ constitute the equivalent PA coefficients of the array. This new reduced complexity learning structure is depicted in Fig. \ref{fig:DPD_simplified}. 

The equivalent array PA model, shown in Fig. \ref{fig:DPD_simplified}, can be identified either by combining the individual PA estimates or more efficiently by means of a combined feedback observation receiver similar to the one considered in \cite{HybridMIMO}, which generates a combination of the PA output signals in the RF domain, from which the equivalent PA can be directly estimated. Such combined observation signal could also be obtained in the digital domain, by first measuring all the individual PA output signals by means of a shared observation receiver, and finally combining the signals in the digital domain. 

\section{Complexity Analysis} 
\label{sec:complexity}
This section presents a quantitative analysis of the computational complexity of the proposed DPD architecture and its comparison against the current state-of-the-art solutions that assume a dedicated DPD block per antenna branch. The number of floating point operations (FLOPs) is used to quantify the complexity. For clarity, we summarize the used notations as follows: $M$ denotes the number of antennas at the transmitter, $Q$ denotes the DPD order, $K$ stands for the number of block-adaptive iterations of the decorrelation-based learning algorithm while $N$ denotes the number of samples for LS estimation, the number of samples per DPD block as well as the number of transmitted symbols. $L$ denotes the length of the RRC filter and $U$ denotes the upsampling factor.

To ensure a fair complexity analysis, we consider the following points and assumptions:

\begin{itemize}
    \item We need to differentiate between DPD learning and actual linearization. The DPD learning is performed only when the PA characteristics or operation point e.g., carrier frequency or bandwidth change, while the linearization is executed continuously along the actual data transmission. 
    \item In current state-of-the-art solutions, the digital precoder works at symbol rate while then $M$ RRC filtering stages are adopted (one per antenna branch). In the proposed solution, there is only a single RRC filtering block, taking place before the spatial precoder, and thus, the spatial precoder works at the upsampled rate.
    \item  A complex multiplication requires 6 FLOPs while a complex addition requires 2 FLOPs. Multiply and accumulate (MAC) operation requires 8 and 2 FLOPs when involving complex and real numbers, respectively. MAC operations are present in most of the mathematical calculations.  Therefore, the complexity resulting from more sophisticated calculations, such as the Cholesky decomposition (CD) when involving complex operations is approximated here by four times the complexity of its real counterpart.
    \item An appropriate orthogonalization stage based on CD is applied to the basis functions such that they acquire better numerical properties and stability during the DPD learning. This orthogonalization is performed per antenna branch in state-of-the-art solutions while only once in the proposed DPD architecture. The least-squares fitting considered for PA identification is also based on the Cholesky decomposition. The complexity of the complex CD is then approximated as $4(N+\frac{Q+1}{6})(\frac{Q+1}{2})^2$ \cite{Matrix_computations}.
    \item The proposed DPD learning requires mimicking the true PAs, the propagation channel and applying the BB precoder in every iteration of the learning algorithm. This involves generating the SNL basis functions and performing the corresponding MAC operations to generate the PA output signals. The complexity resulting from emulating $M$ true PAs during the learning is $MNK(Q+40)$. On the other hand, the channel and precoder require $8MNK$ and $6MNK$ FLOPs, respectively. The reference solution does not require any of these computations, whereas the proposed reduced complexity learning architecture only requires to emulate the effective PA that corresponds to $NK(Q+40)$ FLOPs.
    \item The upsampling and filtering stage is implemented by means of an efficient polyphase interpolator structure.
\end{itemize}
 The learning and full online transmission complexities resulting from applying different DPD schemes are summarized in Table \ref{Tab:Table1}. As an example, we evaluate and show also concrete complexity values for the three addressed architectures considering the following parameters: $Q = 9$, $K = 20$, $N = 100.000$, $M = 32$, $U = 6$, and $L = 32$. During the online transmission, the transmitter performs the following baseband processing: upsampling and filtering, BF generation, DPD filtering and spatial precoding. On the other hand, during the learning, the transmitter performs BF orthogonalization, PA estimation (could be also done offline) and DPD learning. It can be observed that the proposed reduced complexity solution requires 75\% less computations than current state-of-the-art during the actual transmission, and performs 20 times less computations during the DPD learning. Furthermore, the complexity reduction becomes more significant as the number of antennas increases. Therefore, the proposed DPD solution with the reduced complexity learning architecture is a very appealing approach for linearizing large array digital beamforming transmitters. 

\section{Performance Results and Analysis}\label{sec:results}
In this section, the performance of the proposed DPD solution is assessed with extensive simulations. In general, simulation results with the full-complexity and reduced-complexity parameter learning solutions are essentially identical, thus we focus on the latter case for presentation simplicity. As a concrete example, we consider a $16 \times 1$ MISO system, with 16-QAM data modulation, 20 MHz channel bandwidth, and 22\% roll-off in the RRC filtering, while also vary the array size between 4 to 60. We adopt 9-th order clipped memoryless polynomial models for the PA units, obtained through actual RF measurements carried out in the Lund massive MIMO testbed. 
For visualization purposes, we exclude additive channel noise when showing the received signal spectra.
 \begin{figure}[t!]
    \centering
    \includegraphics[width=1\linewidth]{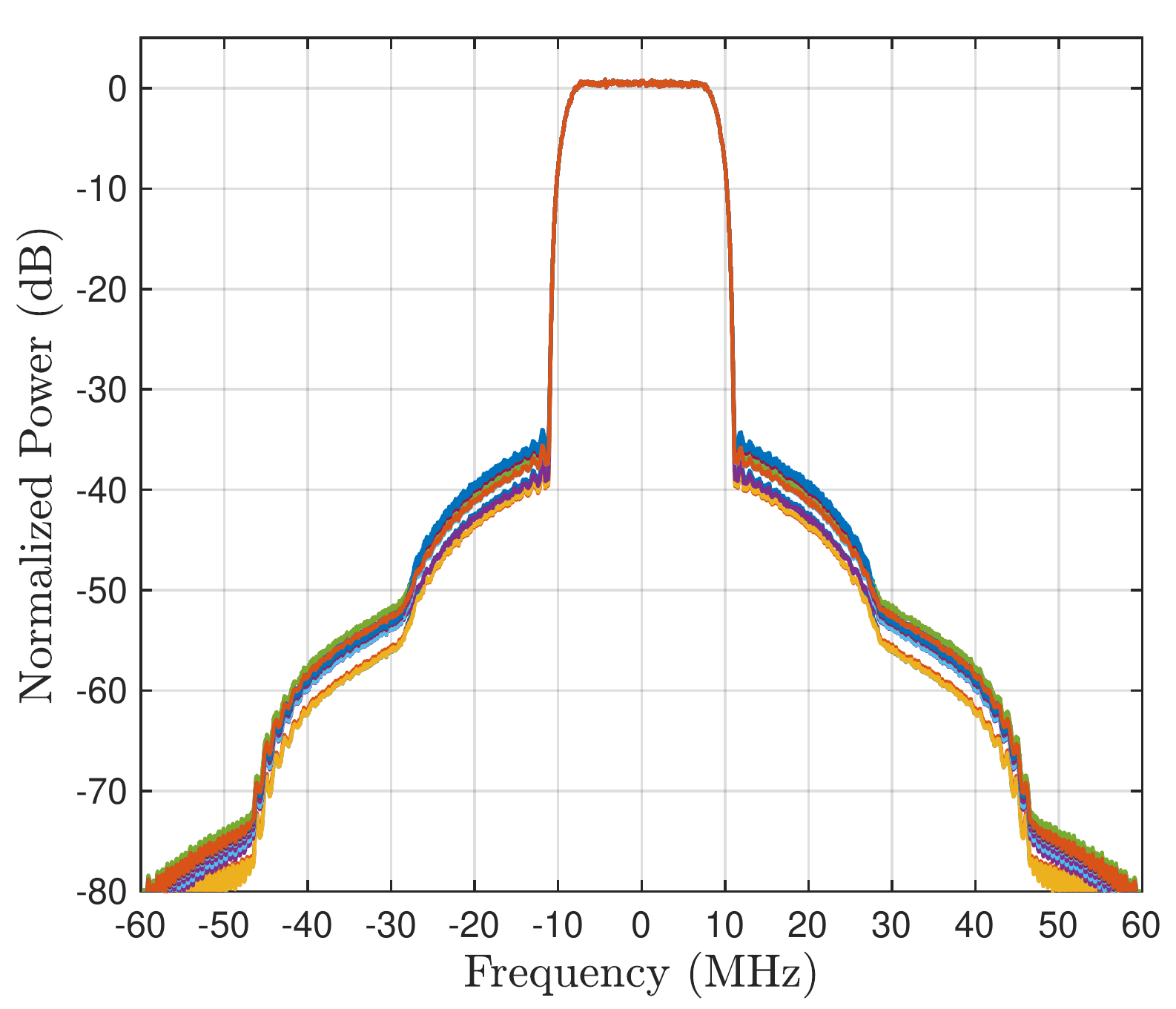}
    \caption{Normalized PA output spectra of 16 different memoryless PA models extracted from a massive MIMO testbed at 120 MHz sample rate. The transmitted waveform is a 20 MHz single carrier waveform with 16-QAM data modulation. The passband power of every PA is normalized to 0 dB.}
    \label{fig:PA_outputs}
\end{figure}
%
%

In the performance evaluations, we consider two fundamental scenarios, defined as follows:

\subsubsection{Scenario A} In this case, the DPD and BB precoder coefficients are first calculated with respect to a specific intended user location and array channel, and high quality linearization at intended receiver is demonstrated. Then, while keeping the DPD and BB precoder coefficients fixed, we randomly draw large amount of victim receiver locations and corresponding array channels, and evaluate the OOB emissions at all receivers as function of the number of transmit antennas. This is then further iterated over different intended RX channel realizations, such that the DPD coefficients and precoder coefficients are recalculated. 

\subsubsection{Scenario B} In the second scenario, the DPD is again learned considering a given location and array channel of the intended user. Then, while keeping the DPD coefficients fixed, the location and the array channel of the intended user are varied and the BB precoder is updated accordingly. Again, the experiment is iterated over different initial intended RX channels used in DPD coefficient calculations.

The first scenario will show that the OOB emissions behave well regardless of the spatial location, while the second scenario will show that the linearization performance at intended RX with fixed DPD coefficients is in practice independent of the channel realization.

 Example output spectra of the 16 different memoryless PA models are depicted in Fig. \ref{fig:PA_outputs}, and we start by evaluating the performance of the proposed DPD solution in the direction of the intended receiver. 
 To this end, we utilize the error vector magnitude (EVM) and ACLR metrics to evaluate the inband quality of the signal and the adjacent channel interference due to spectrum regrowth, respectively \cite{3GPP_BS}, both measured now at intended RX. 
 In general, high order modulations are subject to very strict EVM requirements, down to $1\%$, and the PA nonlinearities can easily violate these limits. The EVM is defined as
\begin{figure}[t!]
    \centering
    \includegraphics[width=1\linewidth]{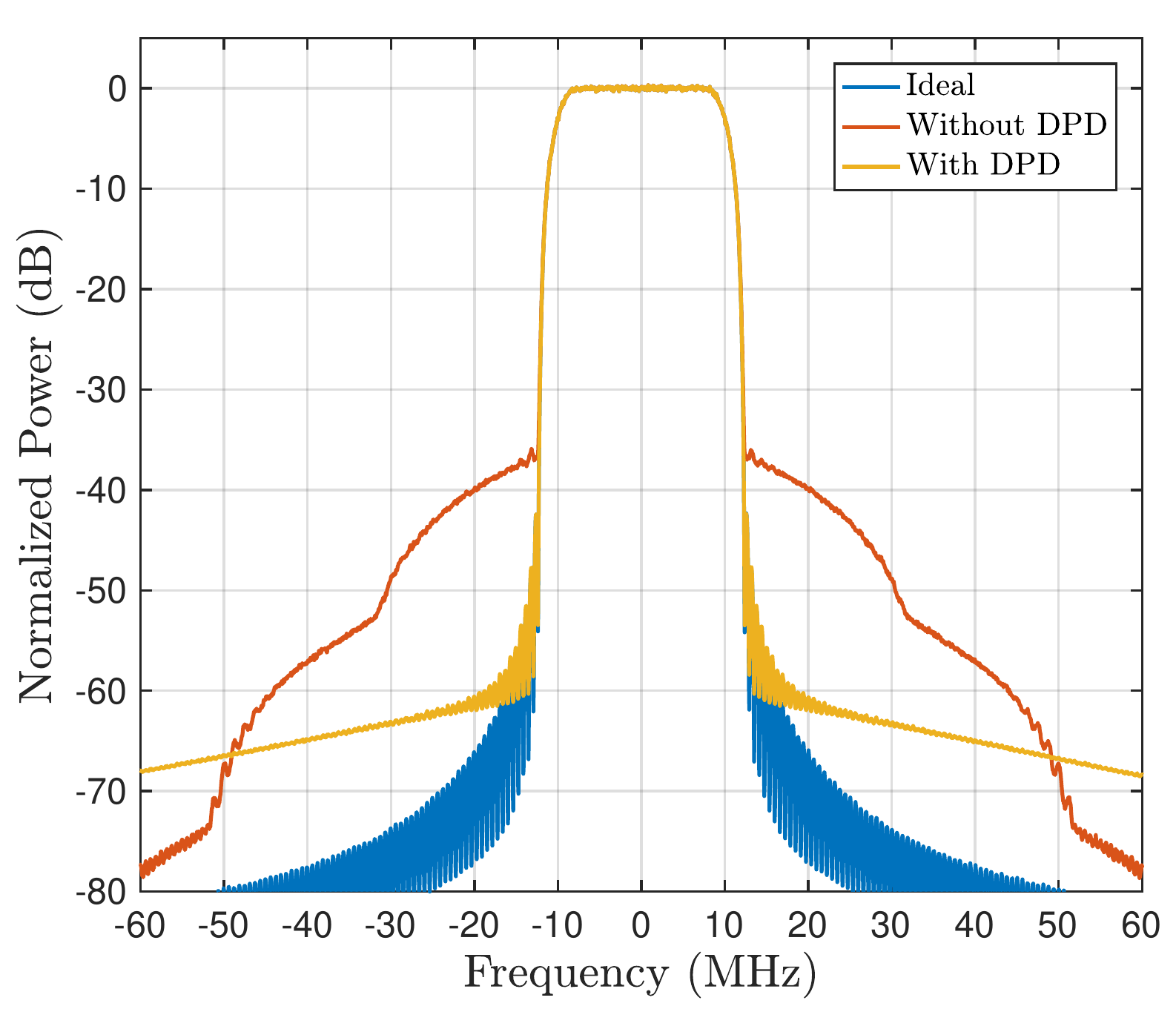}
    \caption{Example normalized spectra of the observed signal at the intended receiver, without and with DPD, with TX array size of 16. EVM and ACLR numbers are 2.08\% and 39.1~dB (without DPD), and 0.64\% and 59.8~dB (with DPD). 
    }
    \label{fig:SU}
\end{figure}

\begin{equation}
EVM_{\%} = \sqrt{P_{error}/P_{ref}} \times 100\%,
\end{equation}
where $P_{error}$ is the power of the error between the ideal symbols and the corresponding symbol rate complex samples at the intended receiver, both normalized to the same average power, while $P_{ref}$ is the reference power of the ideal symbol constellation. On the other hand, the ACLR is defined as the ratio between the powers observed within the intended channel, $P_{intended}$, and within the right or left adjacent channels, $P_{adjacent}$, expressed as
\begin{equation}
ACLR_{dB} = 10 \log_{10} \frac{P_{intended}}{P_{adjacent}}.
\end{equation}
We define the measurement bandwidth at the intended channel as the bandwidth containing $99\%$ of the total observed signal power at the intended receiver. The adjacent channel power has then the same measurement bandwidth, and can be measured either at the intended RX location or other victim receiver locations. Note that both adjacent channels have the same unwanted emission power due to the memoryless modeling.



\begin{figure}[t!]
    \centering
    \includegraphics[width=1\linewidth]{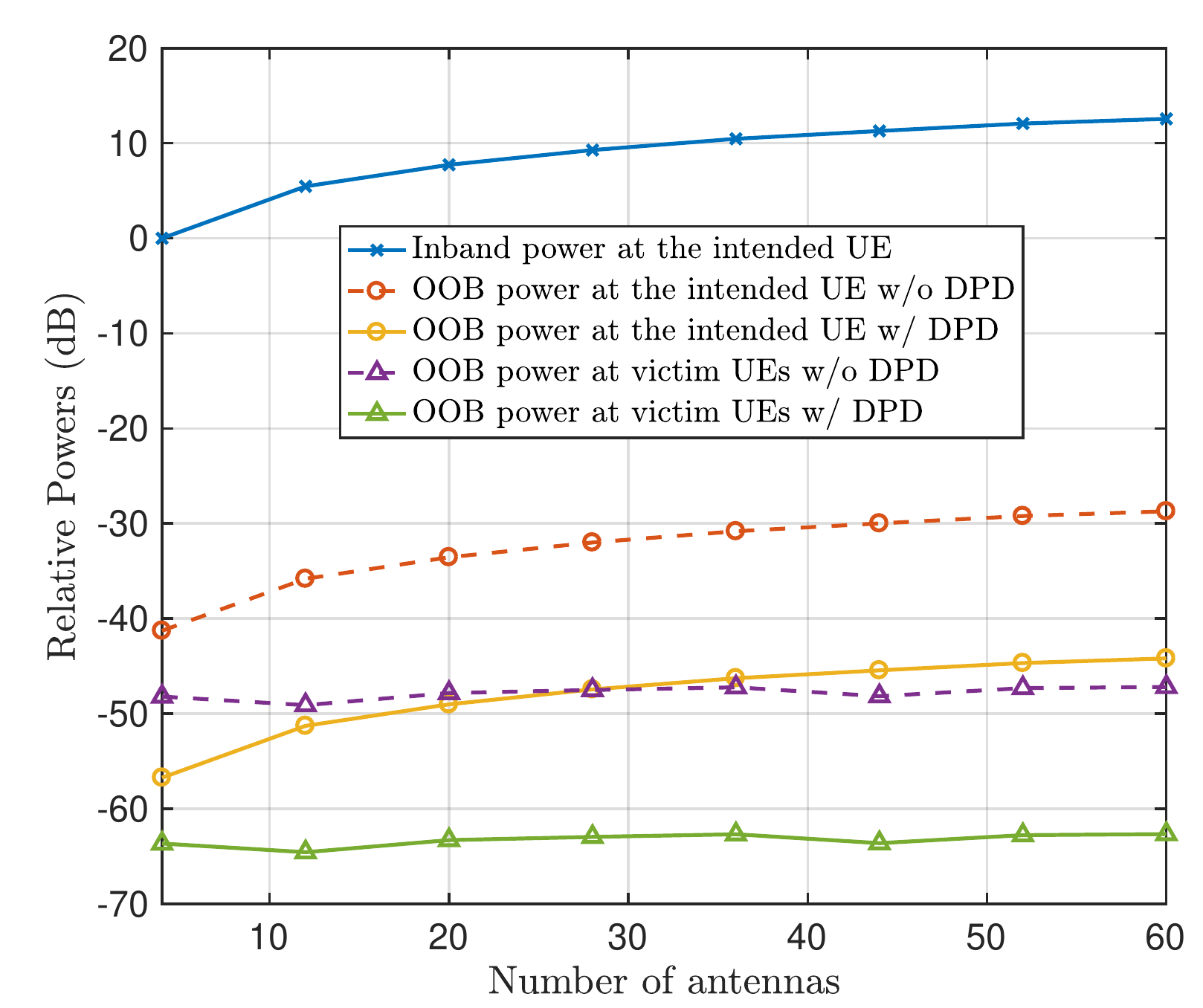}
    \caption{Averaged inband and OOB powers at intended and victim users, with and without DPD, in Scenario A. 
    The powers are normalized such that for the smallest array, the inband power is 0 dBm. 
    }
    \label{fig:array}
\end{figure}
\begin{figure}[t!]
    \centering
    \includegraphics[width=1\linewidth]{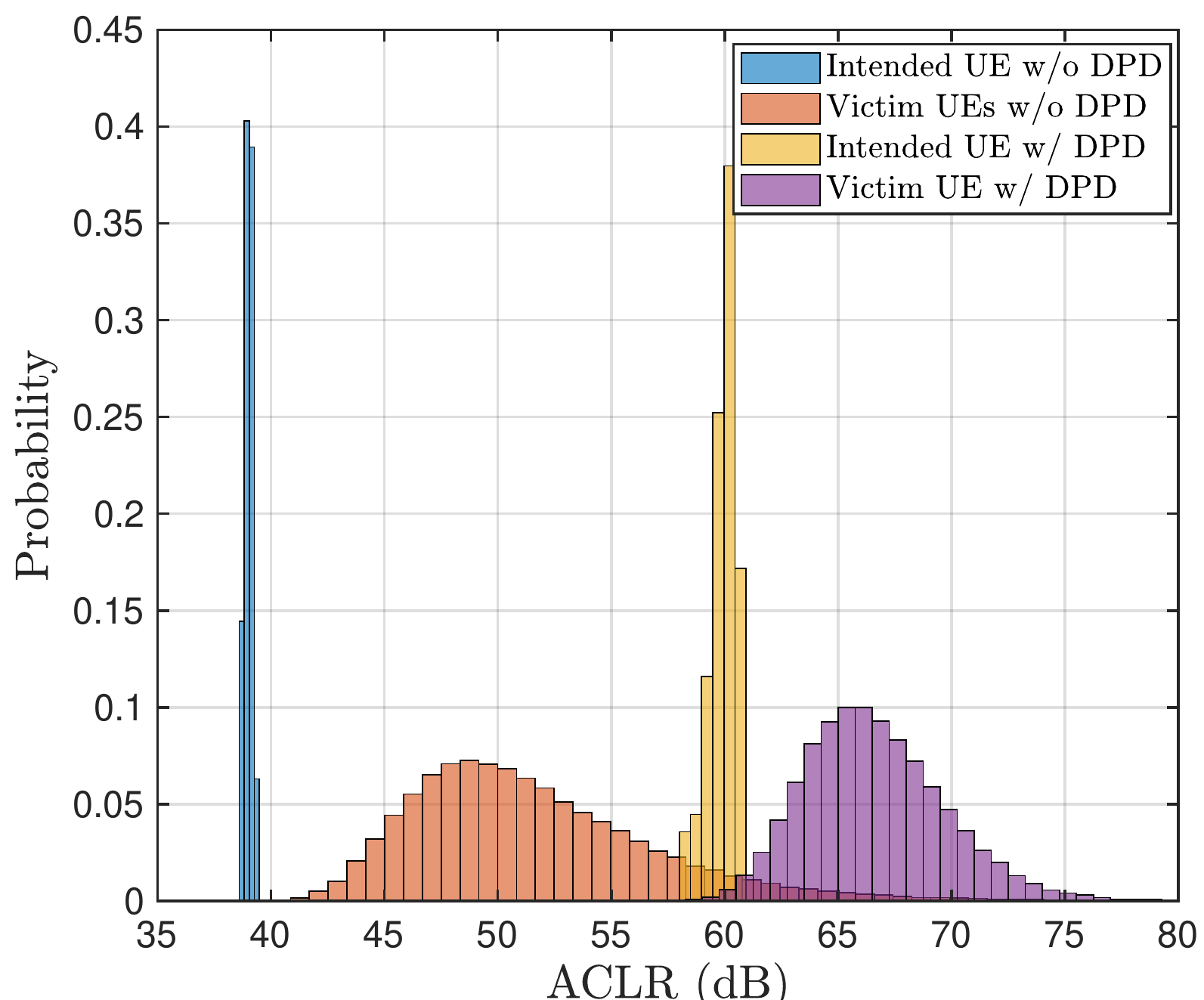}
    \caption{Empirical ACLR distributions at intended and victim receivers, with and without DPD, for an array size of 16 in Scenario A.  
    }
    \label{fig:victims}
\end{figure}
\begin{figure}[h]
    \centering
    \includegraphics[width=1\linewidth]{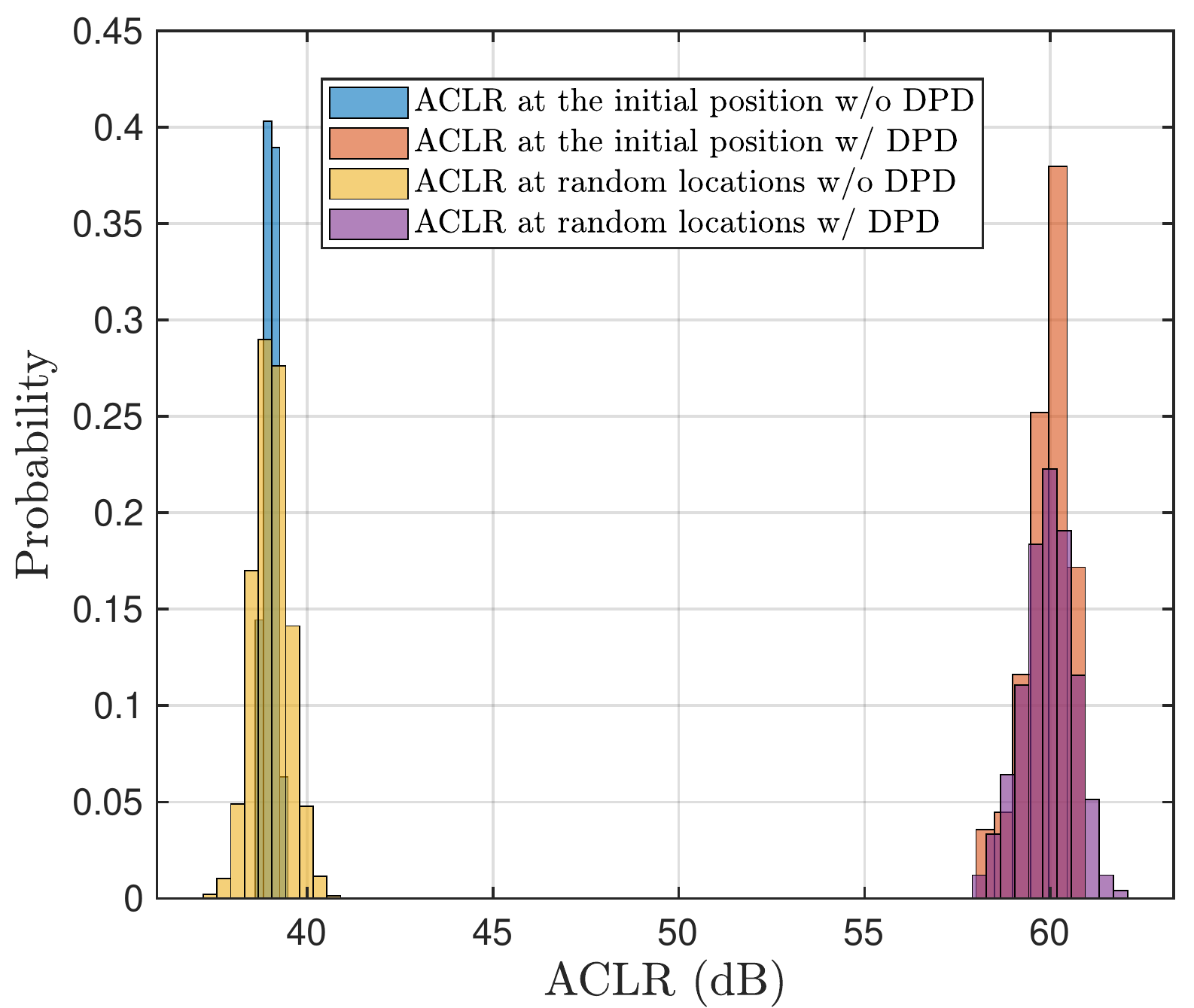}
    \caption{Empirical ACLR distributions at intended receivers, with and without DPD, in Scenario B. 
    }
    \label{fig:dif_locations}
\end{figure}
Example linearization results are shown 
in Fig. \ref{fig:SU}, illustrating how efficiently the out-of-band emissions are reduced at the intended receiver and how the passband EVM is improved. While the previous results demonstrate a snap-shot performance at intended RX with a single channel realization, we next
evaluate how the linearization performance behaves more broadly, when considering different intended RX channel realizations and different victim receivers at random locations while also varying the array size. 
To this end, we consider 200 randomly located victim receivers operating at the adjacent 20 MHz channel and measure the observable OOB powers stemming from our array transmitter, with digital precoder and DPD being calculated for the given fixed location of the intended RX. This is then repeated as a whole for 600 different intended RX locations/channels. The results in terms of the averaged powers are shown in Fig. \ref{fig:array}, for varying number of transmit antennas, where also the passband power level at the intended receiver is shown for reference. 
It can be observed that as the number of antennas increases, the inband and OOB powers towards the intended user increase due to higher beamforming gain, while the proposed DPD method gives a systematic OOB suppression gain of ca. 20 dB. On the other hand, in other spatial directions, the observable OOB powers are more independent of the array size while showing a fairly systematic DPD gain of ca. 15 dB. It can also be noticed that the OOB powers at other victim receiver locations are systematically lower than at the intended RX, which is stemming from less coherent combining of the radiated nonlinear distortion. Overall, the results illustrate that despite the DPD is being learned with a focus on the intended user RX, the OOB emissions are well behaving and systematically suppressed in all victim user locations.

In order to have further insight into the behavior of the unwanted emissions
, we next analyze the empirical distributions of the effective ACLRs 
assuming the baseline case of 16 antennas. The DPD is again learned for a given intended RX channel/location, while then the OOB emissions at intended RX as well as 200 randomly located victim receivers are evaluated. This procedure is then repeated for 600 randomly drawn intended user channels/locations,
to gather the overall OOB emission statistics. The obtained results are depicted in Fig. \ref{fig:victims}. The two more narrow distributions represent the ACLRs without and with DPD at the location of the intended user. Then, the distributions of the ACLRs at the experimented 120.000 randomly located victim receivers, with and without, DPD are shown. The distributions at intended RX evidence systematic linearization independent of the actual channel. It is also important to remark that despite there is some seeming overlap in the victim receiver ACLR distributions without and with DPD, the DPD never makes the emissions worse when experimented down to an individual victim RX level. It can also be observed that there are generally some victim RX cases where the natural spatial suppression for the OOB emissions is already huge, resulting in very low OOB power at victim receiver even without DPD. In general, we can conclude that the proposed DPD system and the array channel together provide good OOB emission suppression for any arbitrary victim receiver, with the average linearization gain being comparable to the one we get at the intended user. 

Lastly, we address the Scenario B, where the DPD is first learned for a given intended receiver location/channel. Then, the location/channel of the intended receiver is randomly varied, with 200 different realizations, and the precoder coefficients are adapted accordingly, while the DPD coefficients remain fixed. This overall setup is then further iterated over 600 different initial locations of the intended receiver. The corresponding empirical distributions of the intended RX ACLRs, without and with DPD, are shown in Fig. \ref{fig:dif_locations}. The results show that the linearization performance of the proposed DPD is very insensitive to the changes in the intended user array channel as long as the PA characteristics stay the same.

\section{Conclusions}\label{sec:conclusion}
In this article, we have introduced a novel DPD architecture for fully digital beamforming transmitters that is far less complex than current state-of-the-art solutions. The proposed solution requires only a single DPD to linearize an arbitrary amount of parallel and mutually different power amplifiers, providing excellent linearization performance. The proposed transmitter architecture exhibits lower complexity, both during the DPD learning and the actual transmission. It also  reduces significantly the HW complexity and implementation cost of the transmitter, since it only requires a single observation receiver for the DPD learning. Extensive simulations considering real PA models were conducted in order to show the efficacy of the proposed architecture and parameter learning solution. Furthermore, out-of-band emissions were evaluated in the spatial domain, and they were shown to remain essentially lower than those at the direction of the intended receiver which is the most harmful direction in terms of nonlinear distortion.
\section*{Acknowledgment}
This work was supported by Tekes, Nokia Bell Labs, Huawei Technologies Finland, RF360, Pulse Finland and Sasken Finland under the 5G TRx project, by the Academy of Finland under the projects 288670 and 301820, and by TUT Graduate School.
\bibliographystyle{IEEEbib}
\bibliography{Ref}

\begin{thebibliography}{10}

\bibitem{Intro_1}
E.~G. Larsson, O.~Edfors, F.~Tufvesson, and T.~L. Marzetta,
\newblock ``{Massive MIMO for next generation wireless systems},''
\newblock {\em IEEE Commun. Mag.}, vol. 52, no. 2, pp. 186--195, February 2014.

\bibitem{Energy_efficiency}
F.~Rusek, D.~Persson, B.~K. Lau, E.~G. Larsson, T.~L. Marzetta, O.~Edfors, and
  F.~Tufvesson,
\newblock ``Scaling up {MIMO}: Opportunities and challenges with very large
  arrays,''
\newblock {\em IEEE Signal Process. Mag.}, vol. 30, no. 1, pp. 40--60, Jan
  2013.

\bibitem{5Green}
M.~Olsson, C.~Cavdar, P.~Frenger, S.~Tombaz, D.~Sabella, and R.~Jantti,
\newblock ``5{G}reen: Towards green {5G} mobile networks,''
\newblock in {\em 2013 IEEE 9th Int. Conf. on Wireless and Mobile Computing,
  Networking and Commun. (WiMob)}, Oct 2013, pp. 212--216.

\bibitem{GreenComm}
L.~Guan and A.~Zhu,
\newblock ``Green communications: Digital predistortion for wideband {RF} power
  amplifiers,''
\newblock {\em IEEE Microw. Mag.}, vol. 15, no. 7, pp. 84--89, Dec. 2014.

\bibitem{Intro_6}
J.~Shen, S.~Suyama, T.~Obara, and Y.~Okumura,
\newblock ``{Requirements of power amplifier on super high bit rate massive
  MIMO OFDM transmission using higher frequency bands},''
\newblock in {\em 2014 IEEE Globecom Workshops (GC Wkshps)}, Dec 2014, pp.
  433--437.

\bibitem{OOB_Mollen}
C.~Moll\'{e}n, E.~G. Larsson, U.~Gustavsson, T.~Eriksson, and R.~W. Heath,
\newblock ``{Out-of-Band Radiation from Large Antenna Arrays},''
\newblock {\em IEEE Commun. Mag.}, vol. 56, no. 4, pp. 196--203, April 2018.

\bibitem{DPD_MM_1}
M.~Abdelaziz, L.~Anttila, and M.~Valkama,
\newblock ``{Reduced-complexity digital predistortion for massive MIMO},''
\newblock in {\em 2017 {IEEE} ICASSP}, March 2017, pp. 6478--6482.

\bibitem{HybridMIMO}
M.~Abdelaziz, L.~Anttila, A.~Brihuega, F.~Tufvesson, and M.~Valkama,
\newblock ``Digital predistortion for hybrid {MIMO} transmitters,''
\newblock {\em {IEEE} J. Sel. Topics Signal Process.}, vol. 12, no. 3, pp.
  445--454, June 2018.

\bibitem{DPD_MM_3}
H.~Yan and D.~Cabric,
\newblock ``{Digital predistortion for hybrid precoding architecture in
  millimeter-wave massive MIMO systems},''
\newblock in {\em 2017 {IEEE} ICASSP}, March 2017, pp. 3479--3483.

\bibitem{DPD_MM_2}
L.~Liu, W.~Chen, L.~Ma, and H.~Sun,
\newblock ``{Single-PA-feedback digital predistortion for beamforming MIMO
  transmitter},''
\newblock in {\em 2016 IEEE ICMMT}, June 2016, vol.~2, pp. 573--575.

\bibitem{DPD_MM_4}
S.~Lee, M.~Kim, Y.~Sirl, E.~R. Jeong, S.~Hong, S.~Kim, and Y.~H. Lee,
\newblock ``{Digital predistortion for power amplifiers in hybrid MIMO systems
  with antenna subarrays},''
\newblock in {\em 2015 IEEE VTC}, May 2015, pp. 1--5.

\bibitem{DPD_HMIMO}
X.~Liu, Q.~Zhang, W.~Chen, H.~Feng, L.~Chen, F.~M. Ghannouchi, and Z.~Feng,
\newblock ``Beam-oriented digital predistortion for {5G} massive {MIMO} hybrid
  beamforming transmitters,''
\newblock {\em {IEEE} Trans. Microw. Theory Tech.}, vol. 66, no. 7, pp.
  3419--3432, July 2018.

\bibitem{Dec_Based}
M.~Abdelaziz, L.~Anttila, C.~Tarver, K.~Li, J.~R. Cavallaro, and M.~Valkama,
\newblock ``Low-complexity subband digital predistortion for spurious emission
  suppression in noncontiguous spectrum access,''
\newblock {\em {IEEE} Trans. Microw. Theory Tech.}, vol. 64, no. 11, pp.
  3501--3517, Nov 2016.

\bibitem{Concurrent_abdelaziz}
M.~Abdelaziz, L.~Anttila, A.~Kiayani, and M.~Valkama,
\newblock ``Decorrelation-based concurrent digital predistortion with a single
  feedback path,''
\newblock {\em {IEEE} Trans. Microw. Theory Tech.}, vol. 66, no. 1, pp.
  280--293, Jan 2018.

\bibitem{Matrix_computations}
Gene~H. Golub and Charles~F. Van~Loan,
\newblock {\em Matrix Computations (3rd Ed.)},
\newblock Johns Hopkins University Press, Baltimore, MD, USA, 1996.

\bibitem{3GPP_BS}
{\em {LTE {E}volved {U}niversal {T}errestrial {R}adio {A}ccess {(E-UTRA)}
  {B}ase {S}tation {(BS)} radio transmission and reception, 3GPP TS 36.104
  V11.8.2 (Release 11)}}, April 2014.

\end{thebibliography}

\end{document}